\documentclass[AMA,LATO1COL]{WileyNJD-v2}

\articletype{Article Type}%

\received{22 May 2021}
\revised{6 June 2021}
\accepted{6 June 2021}

\usepackage{capt-of}

\usepackage{amsmath,amssymb,amsfonts}
\usepackage{graphicx}
\usepackage{textcomp}
\usepackage{cleveref}
\usepackage{xcolor}
\usepackage{todonotes}
\usepackage[utf8]{inputenc}
\usepackage[official]{eurosym}
\usepackage{soul}
\usepackage{subcaption}
\usepackage{graphicx} 
\usepackage{tabularx}
\usepackage{amssymb}
\usepackage{pifont}
\newcommand{\cmark}{\ding{51}}%
\newcommand{\xmark}{\ding{55}}%

\usepackage{todonotes}

\usepackage{amssymb}
\usepackage{pifont}

\newboolean{showcomments}
\setboolean{showcomments}{true} 
\ifthenelse{\boolean{showcomments}}
{\newcommand{\nb}[2]{
  \fcolorbox{black}{yellow}{\bfseries\sffamily\scriptsize#1}
  {\sf$\blacktriangleright$\textit{#2}$\blacktriangleleft$}
 }
 
}
{\newcommand{\nb}[2]{}
 
}

\newcommand\editcolor{black}

\begin{document}

\title{Internet-of-Things Architectures for Secure Cyber-Physical Spaces: the VISOR Experience Report}


\author [1]{*Jessica De Pascale}

\author[2]{Giuseppe Cascavilla}

\author[1]{Mirella Sangiovanni}

\author[2]{Damian A.~Tamburri}

\author[1]{Willem-Jan van den Heuvel}

\address[1]{\orgdiv{University of Tilburg \& JADS}, \orgaddress{\state{Netherland}, \country{'s-Hertogenbosch}}}

\address[2]{\orgdiv{Eindhoven University of Technology \& JADS}, \orgaddress{\state{Netherland}, \country{'s-Hertogenbosch}}}

\abstract[Summary]{Internet of things (IoT) technologies \textcolor{\editcolor}{are becoming a more and more widespread} part of civilian life in common urban spaces, which are rapidly turning into cyber-physical spaces. Simultaneously, the fear of terrorism and crime in such public spaces is ever-increasing. Due to the resulting increased demand for security, video-based IoT surveillance systems have become an important area for research.
Considering the large number of devices involved in the illicit recognition task, we conducted a field study in a Dutch Easter music festival in a national interest project called VISOR to select the most appropriate device configuration in terms of performance and results. We iteratively architected solutions for the security of cyber-physical spaces using IoT devices. We tested the performance of multiple federated devices encompassing drones, closed-circuit television, smart phone cameras, and smart glasses to detect real-case scenarios of potentially malicious activities such as mosh-pits and pick-pocketing. Our results pave the way to select optimal IoT architecture configurations---i.e., a mix of CCTV, drones, smart glasses, and camera phones in our case---to make safer cyber-physical spaces' a reality.}


\maketitle

\section{Introduction} \label{sec:introduction}

We are living in a world that is moving towards a reality where public spaces are subject to the operation of multiple interconnected devices (e.g., steady- and speed-cameras) as part of a phenomenon collectively referred to as Internet-of-Things (IoT)~\cite{GeorgakopoulosJ16,kranz2009embedded}. 
On the one hand, such devices empower collecting big data through multiple sensors that surveillance systems can exploit to monitor and secure physical spaces.
On the other hand, the interplay of different devices makes up complex software architectures~\cite{MisicAM18} that could hinder stakeholders in architecting and using surveillance technologies in the so-called cyber-physical spaces (CPS)~\cite{TsigkanosKG16}. 
In this context, motivated by the increasing demand for security in public space, the problem has arisen of finding the most appropriate assistive security technology featuring IoT \cite{guimaraes2018proposal}. 
With the term ``assistive security'', we indicate a novel conceptualization around applied IoT computing and intended as any device that can provide technological or other support to law enforcement or public safety \cite{CASCAVILLA2021102258}.

To provide a compass to cope with this need for direction, we report our experience in the scope of a national interest project in The Netherlands called VISOR. Originally intended to pave the way towards developing, exploring, and validating a dynamic, software-enabled, and data-driven contingency planning~\cite{clark2010contingency}, VISOR resolves around the development of a framework to assist stakeholders in architecting, maintaining, and using IoT-enabled surveillance technologies in events supported by cyber-physical technology and the software backing it. 

The VISOR project has been thought in collaboration with the Noord-Brabants Dutch Police and municipality stakeholders and is meant to focus on providing a platform for the monitoring and detection of unsafe/unsecured behaviour in large crowds (e.g., think of mosh-pits or pick-pocketing) \textcolor{\editcolor}{using IoT devices.}


\textcolor{\editcolor}{Motivated by the stakeholders' demand for the protection and security of public spaces with the most appropriate IoT devices and the need for an architecture to handle the interconnection among them, } we aim to answer the master research question (Main RQ) and sub-research questions (SRQs) listed below.

\begin{center}
    \textit{\textbf{Main RQ:} How can the most appropriate IoT architecture to secure cyber-physical spaces based on experienced stakeholders' concerns be designed?}    
\end{center}

\begin{itemize}
    \item [\textbf{SRQ$1$}] What mapping exists between available non-functional metrics and required stakeholder concern?
    
    
    \item [\textbf{SRQ$2$}] What mapping exists between available non-functional metrics and IoT devices?
    
    \item [\textbf{SRQ$3$}] What experience do users perceive when using governed IoT architectures?
\end{itemize}

\textcolor{\editcolor}{More specifically, the stakeholders involved in the VISOR project---a national project featuring the Dutch Police, the HSG company and other interested parties such as the Municipality of Schijndel---were demanding a reusable reference architecture designed to integrate already available IoT devices requiring for data-flows across such devises to be supported in combination with appropriate analytics.}

To address the aforementioned questions and research goals, we instantiated an exploratory field study at Paaspop\footnote{namely, the Holland Security Group (HSG)\url{https://www.hollandsecuritygroup.com/}}, a three-day event during the Easter weekend, which traditionally kicks off the festival season in The Netherlands. 
This field study is grounded on the Paaspop event for three main reasons: (i) counting 15 stages, more than 175 bands, and more than 90 thousand people~\cite{paaspop}, the festival represents a concrete example of an environment where critical accidents can originate; (ii) the possibility of comparing it to similar events such as concerts, markets, parades, where safety against dangerous situations or terroristic attacks is of paramount importance; (iii) the complexity of the festival requires different technologies and approaches. 

Our contributions in this paper can be summarized as follows:

\begin{enumerate}
    \item \textbf{Design Principles}. We found relevant relations between the concerns raised by stakeholders and non-functional metrics exhibited by the IoT technology under study - these are valuable in further experiments and can be used as design principles in the same context.
    \item \textbf{Platform Architecture}. We elaborated a tested-true architecture to protect semi-public cyber-physical spaces featuring IoT - this is a valuable starting point for further case studies on the matter or even replications of experiences such as the one reported in this paper.
    \item \textbf{Lessons Learned}. We report the pros and cons and lessons learned from each device's usage as part of our field study and the monitoring process within it - these would be essential to recreate or reuse the same proposed architecture in action.
    \item \textbf{IoT Experience Report}. We report critical security guards' feedback on the combined usage for the proposed architecture; for example, we discovered several challenges with using detection via IoT smartglasses. 
\end{enumerate}

We concluded that the proposed framework evaluated in the Paaspop field study helps to select the most optimal architecture configuration---i.e., a mix of CCTV, drones, smart glasses, and camera phones in our case---that yields optimized cyber-physical spaces' security according to the involved security offices and Law-Enforcement Agencies (LEAs) involved in our study. 

\textbf{Structure of the Article}. The rest of this paper is structured as follows. In \cref{sec:problem} we present the problem statement and the Paaspop study with the stakeholders involved (Sec.~\ref{sec:VISOR}-\ref{sec:stakeholders}). Then, \cref{sec:research_design} focuses on the research design, showing the infrastructure and the process of the study conducted. Further on, in \cref{sec:exp_setup} the experimental setup has been described, analyzing the research methods, approaches, and metrics involved in each research question. \Cref{sec:dataset} provides the dataset overview and describes the correlation between parameters and the features shown. 
In \Cref{sec:results} the results of each research question have been shown, highlighting the experimental process behind the Paaspop field study. 
\Cref{sec:ttv} raises the threats to validity raised in this work. 
Finally, sections \ref{sec:discussion} and \ref{sec:conclusion} discuss the results and outline conclusions and future work.
\section{Background, Problem Definition and Context} \label{sec:problem}
Cyber-physical systems (CPS) are systems capable of making decisions independently over a physical scenario, e.g., using some form of artificial intelligence or rule-based decision-making system \cite{UllrichW15}. From an architectural perspective, a CPS is the combination of different software systems with different nature, often explicitly connected to the technology domain referred to as IoT, or micro-hardware with the common aim to control, monitor, and operate a physical process and---through continuous feedback both from that process and the running software---be able to adapt to the different conditions of both process and system at runtime \cite{TsigkanosKG16}. 

\subsection{Background and Related Work}

While previous studies have identified and discussed the need for more reference and automated support in such architectures---especially so in the context of complex and IoT-supported scenarios (e.g., think of a smart city or even simply a smart event)~\cite{kranz2009embedded}---previous work in IoT has had a relatively limited lens of analysis, mainly focusing on guaranteeing software architecture properties \cite{tan2018software}, working towards edge/fog computing \cite{TomovicYMR17} and similar computer-specific aspects of the same technology. 


\textcolor{\editcolor}{A study performed by Rodda et al. \cite{rodda2020suspect} shows an architecture for suspicious visitor monitoring, using camera footage, event detection, and tracking systems featuring a Raspberry~Pi architecture. The authors proposed a methodology to gather and process visitors' images with the intent of notifying the event security department if a suspicious visitor is detected.}

\textcolor{\editcolor}{Furthermore, Bazhenov et al. \cite{bazhenov2019event} present an edge-centric microservice-oriented system addressing the problem of \emph{en-field} data processing with multiple video streams.}

\textcolor{\editcolor}{From an architectural perspective, an urban IoT architecture focused on the needs of cities and their stakeholders have been described by Luckner et al. \cite{luckner2020iot}. They developed a platform for intelligent transport planning using several data sources (e.g., public transport GPS, timetables, Twitter feeds).}

\textcolor{\editcolor}{Each of these works deals with the CSP protection problem differently. Conversely, this work is a preliminary analysis focusing on developing an architecture to recommend the best IoT device for the concerns raised by the stakeholders. It presents an architecture that works with multiple IoT devices and multiple sources (e.g., GPS, video streams, pictures) in several social event scenarios, such as crowd social events, festivals, and social gatherings.}






\subsection{Problem Statement and Context Definition}

Remarkably, at the point in which this study was enacted, we had to report that an experience-grounded reference architecture for IoT-enabled, safety-specific, cyber-physical systems was not proposed yet, other than simplistic examples~\cite{muccini2019iot} which were very much speculative concerning the capacities of such proposals to perform in real-life---rather than \emph{realistic}---scenarios.

Conversely, in the scope of our scenario within the Paaspop VISOR experimentation, physical spaces and software components came together with IoT technology and emerged as strictly connected to guarantee safety and security to all the participants; this led the same scenario to result into a \emph{cyber-physical space}, by definition \cite{TsigkanosKG16}. Within such a space and during the social events intended for it, heavyweight crowd management techniques (e.g., ground personnel for crowd-control and steering) become crucial to ensure the safety, planning actions aimed to increase the level of security of any given place across space. Multiple devices can be involved in this process \cite{sharma2018review} (e.g., crowd data acquisition techniques based on Vision, Wireless/Radio-Frequency (RF), and Web/Social-media data mining technologies). At the same time, a strategic plan---with a robust IoT software architecture behind it---can be used to make the difference between the safety of a place or its vulnerability to all sorts of nasty crowd scenarios, including terrorist attacks.

The problem we aim to address in this study is the choice of the best recommended IoT devices used in public spaces' monitoring. It is crucial to find which combination of IoT devices and software support ---or what we define here a complex \emph{IoT software architecture} for \emph{Secure Cyber-Physical Spaces}--- is most appropriate for the stakeholders and domain involved and for which specific crowd protection scenario it best addresses. 

\subsection{The VISOR project in pills} \label{sec:VISOR}

As aforementioned, VISOR is a project born in collaboration with the Dutch government, a Law Enforcement Agency (LEA) as the Dutch police,
and the Paaspop festival organization team during spring 2019. The project aimed to provide new methods, techniques, methodologies, and devices to enhance the video-surveillance capabilities for the stakeholders involved, with experimented and tested-true IoT software architectures. More in detail, the goal of the project was to:
1) Provide an overview about technologies, IoT devices, and sensors to use in dynamics social events environment, where an illicit behaviour can be brought to fruition.
2) Prepare context-specific detection automation to recognize deviant crowd activities like mosh-pit and pick-pocketing.
3) Elaborate on the feasibility of the proposed solutions to enhance the safety and security of public spaces while preserving the privacy of involved users.  
 

To provide some numbers for the Paaspop event (which spans 3 full days around Easter every year in The Netherlands), we have to consider that there are up to 15 different stages during the festival with more than 175 bands, including a wide range of musical and theater genres, visited by around 40K visitors per year, with a cap to 44343 visitors on the year we enacted the VISOR experimentation (i.e., 2019 \footnote{\url{https://nltimes.nl/2019/04/22/record-number-visitors-paaspop-festival}}). Next to its broad music offerings, Paaspop also boasts many other activities, including a barbershop, hot tubs, a camping shop, a coffee lounge, a retro seventies-eighties roller-skate track, an old school Arcade hall, and an actual cinema. Multi-day ticket holders typically reside at the luxurious camping site of the festival. The camping site boasts tents and festival caravans, huts, and glamping-like festival tents such as Wigwags.



\subsection{Gathering the VISOR Stakeholders' Perspective: a Focus-Group Approach} \label{sec:stakeholders}



This section presents the organization behind the VISOR project and the research activities exercised during the event directly on the Paaspop field, starting from an overview of the preliminary preparations.

\subsubsection{VISOR operations at Paaspop: Preliminaries}
First, four months before the main Paaspop event, the authors could meet the organizers to collect information regarding the festival in 5 distinctive half-day workshops. The main research activity during these workshops with the executive team focused on three main aspects: 
\begin{enumerate}
    \item \textbf{Collection of requirements}. We aimed to gather the stakeholders' expectations which they have from the outcome of the project.
    \item \textbf{Type of technology available on the grounds}. During this process, we collected information regarding the devices available during the event to protect spaces. 
    \item \textbf{Network system and software support}. We collected information about the network system at our disposal available during the event and the software already being used during the same event. 
\end{enumerate}

This first part of collecting information resulted in data to use for instrumenting our VISOR project in action. More specifically, the organizers highlighted some significant critical deviant behaviours: 
\begin{itemize}
    \item Illicit Drug Trade; 
    \item pick-pocketing in different areas of the event (camping side where there are the tents, zigzagging into the crowd, benefit of lower attention of drunk participants);
    \item formation of moshpits during specific musical moments;
    \item possible crowd attack scenarios (terrorists, armed people, brawls);
    \item illegal entry at a boundary gate or in vulnerable areas around the perimeter;
    \item sentry of a specific individual in the crowd (e.g., a crowd-surfing musician);
\end{itemize}

It is relevant to specify that the VISOR project is a preliminary test to explore, evaluate and improve the protection of spaces and scenarios as relevant instances for crowd protection. In this regard, we focused on the most critical scenarios prone to be readily software-supported with the IoT technology at hand, namely, the pick-pocketing and suspective behaviour patterns among the crowd, mosh-pits under the main stage, and specific user recognition using the VISOR logo and personalized t-shirts. 

The second part of the requirements-gathering exercise focused on the type of technology available at the event. After analyzing the received information, we concluded that CCTV was the primary technology available during the event. All the CCTV were remotely managed, and one single office was in charge of surveillance. Therefore, it has been essential to have a centralized office for CCTV to have the possibility to plug our machine learner for the crowd behaviour surveillance.  

Lastly, we could use a private WiFi channel connected to the central infrastructure regarding the network system. Hence, all our devices have been connected to the main WiFi channel of the event and use the connection to retrieve data regarding the event. 

\subsubsection{VISOR operations at Paaspop: Days of the Event}
The second part of the project operation was the organization of VISOR research activities directly on the Paaspop field, with multiple active groups and different experts involved. 
In total, we built a team of ten responsible experts plus nine students for the experiments. The ten responsible experts were divided into four main teams responsible for different experimentation areas:
\begin{enumerate}
    \item A group responsible for testing the machine learning algorithms;
    \item A group responsible for drones;
    \item A group responsible for the organization of the LEAs that were using smart glasses and camera phones;
    \item The last group responsible of the organization of the students to appear as a ``criminals'' or special guests among the crowd.
\end{enumerate}


Among these four teams, we organized a meeting every hour to keep all the teams updated on the different parts of the VISOR project and share information regarding the ongoing activities and eventually related problems and concerns. This latter activity was structured according to the guidelines of hackathon execution \cite{gama2018hackathons} and featured a Scrum-like agile methods approach \cite{abrahamsson2017agile} for any software component required (a total of 7 software components were eventually developed in the scope of the 3-day event).

\subsubsection{VISOR Research Operations: Relevant Stakeholders and Setting}
\begin{table}[ht!]
\centering
\footnotesize
\renewcommand{\arraystretch}{1.5}
\begin{tabular}{p{2cm}p{2cm}p{2cm}p{6.5cm}}
\toprule
 \textbf{Stakeholders} & \textbf{Role} & \textbf{Background} & \textbf{Concerns} \\
\toprule
\multirow{4}{*}{3 $\times$ LEAs} & \multirow{4}{*}{Support} & \multirow{4}{*}{Video surveillance} & \textbf{(1)} The number of people in a social event could be so high that it is difficult to identify fraudulent criminal scenario \\ 
 & & & \textbf{(2)} Large space to monitor. Difficult to identify hot-spots \\
 & & & \textbf{(3)} Lack of insights on hot-spots and anomalous behaviour \\ 
 & & & \textbf{(4)} Frequent cases of pick-pocketing in social events \\

5 $\times$ Paaspop festival organization & Main head of the project & Private security department, hired with the scope to ensure the safety of the event & \textbf{(5)} Complexity and difficulty to provide a real time tracking to the security it staff that help them in the security process; leading to many delays and misunderstanding in the communication process between central monitoring room and security staff. \\
\bottomrule
\end{tabular}

\caption{Description of the stakeholders involved during the Paaspop 2019 event and their relative concerns.}
\label{tab:stakeholders}
\end{table}

The stakeholders involved in this study have been introduced in the previous section. However, in \Cref{tab:stakeholders} we provide a more detailed description for each of the roles.
Each stakeholder is described with three categories:

\begin{itemize}
    \item [-] \textbf{Role}: there are two types of roles, support and main. A support stakeholder principally provides information and bureaucratic support but is not directly involved in the experimentation process (e.g., staff, security guards, developers, researchers). Instead, a stakeholder with the main role provides direct support on the field, helping in loco or remotely.
    \item [-] \textbf{Background}: this category helps to understand the expertise of the stakeholder.
    \item [-] \textbf{Concerns}: each stakeholder has some concerns for public security. This is the main reason for their involvement in the Paaspop case study and, from their concerns, we created the case scenarios described in \Cref{tab:case_scenarios}. 
\end{itemize}

From the crowd and CPS monitoring perspective, we identified two macro-categories: (a) the LEAs needed to guarantee the \textit{security} of the people during the event through security strategies, risk assessment, threat and vulnerability protection; (b) the Paaspop organization team has the scope to ensure the \textit{safety} of the event taking to anticipate what can happen during the event and the possible consequences, be always ready to act, introduce fail-safe mechanisms (including software-specific components to the overall CPS architecture) so that the entire organization functions even if unforeseen incidents occur.


As aforementioned, the concerns in \Cref{tab:stakeholders} have been collected in the different meetings, where each stakeholder talked about their experience and the concerns involved. Among these concerns, we extracted five general concerns, described in detail in \Cref{tab:stakeholders}: four of them was raised by the LEA and they revolve around topics as crowd management (1), hot spots (2), anomalous behaviour (3) and pick-pocketing (4). The last concern (5), raised by the Paaspop organization, talks about real-time implementation to provide fast help in the security process.

From these concerns, we outlined eight scenarios (a detailed description is provided in \Cref{tab:case_scenarios}) grouped as follows:


%

\begin{itemize}
    \item [1)] \textbf {Security enforcer group} (scenario 1, 2, 3, 5, 7, 8). We decided to include technologies like smart glasses, camera phones, CCTV, and the cloud for this scenario. The idea of interconnecting all the aforementioned technologies is to realize a stream of data analyzed by Machine Learning (ML) algorithms or directly by the Dutch police and extract telemetry information. The benefit of telemetry is the ability to monitor the state of a specif environment, like it could be the stage of a concert or the square of public space. Hence, the information is streamed directly to a centralized platform, ready to be analyzed.
    
    \item [2)] \textbf {Safety enforcer group} (scenario 4, 6). The technology involved in this scenario is smart glasses, camera phones, CCTV, and the cloud. However, in this scenario, we aim to use the previously mentioned technology to predict the evolution of a scenario based on the data streamed by the sensors. The idea is to predict illicit crowd schemes like zigzagging to find people to pick-pocketing, mosh-pits, and more anomalous crowd behaviours.
    
    \item [3)] \textbf{Maintainer group} (scenario 9). The technology involved in this scenario is smart glasses, camera phones, CCTV, and the cloud. In this scenario, we aim to control the energy consumption of those devices that use batteries, hence to be able to intervene on time with backup batteries to continue the stream of data.    
    
    \item [4)] \textbf{Event architect group} (scenario 10, 11). The technology involved in this scenario is smart glasses, camera phones, CCTV, and the cloud. The goal of the scenario is to provide the event architect information regarding the areas of the event itself, spots around the event that require the executive team's attention and intervention. 
\end{itemize}

\subsubsection{Building the VISOR Platform: Relevant Concerns}
\begin{table*}[t]
\centering
\footnotesize
\renewcommand{\arraystretch}{1.5}
\begin{tabularx}{.85\linewidth}{ccX}
\toprule
\textbf{Case Scenario ID} & \textbf{Concern ID} & \textbf{Description} \\
\toprule
CS-1 & 1 & As a security-enforcer I want to be able to get object telemetry from what I am observing, with as many details as possible on the level of social, organizational or other hazard.\\
\midrule
CS-2 & 2 & As a security-enforcer I want to receive telemetry from remote that leads me to point of interest in case of emergency and other special scenarios.\\
\midrule
CS-3 & 3 & As a security enforcer I want to receive help and possibly long-range video feed from remote concerning the hazard area I am currently targeting with my VISOR smartglass.\\
\midrule
CS-4 & 5 & As a safety-enforcer I want to be accompanied by remote video-feed to help me track the evolution of a safety-critical scenario.\\
\midrule
CS-5 & 2 & As a security-enforcer I want to be informed with predictive telemetry and visualization maps concerning hotspots.\\
\midrule
CS-6 & 5 & As a safety-enforcer I want to be able to manage the generation and breaking of special crowd behavior (e.g., moshpits, anomalous trajectories, anomalous group-formation) with certain multi-technological localization (e.g., by triangulation).\\
\midrule
CS-7 & 3 & As a security-enforcer I want to be directed from remote on the trajectory to better reach help in pre-specified scenarios.\\
\midrule
CS-8 & 4 & As a security-enforcer I want to be informed about anomalous trajectories nearby sensitive areas of my interest.\\
\bottomrule
\end{tabularx}
\caption{List of the case scenarios (CS) dealt in the PaasPop event.}
\label{tab:case_scenarios}
\end{table*}

Stemming from the research above setup, several concerns emerged in the scope of the focus groups. Said concerns are reported in \Cref{tab:case_scenarios}. On the table, each scenario (ID in Column 1) is mapped to a specific concern (ID in column 2) and reflects a concern raised by the stakeholders in the form of an agile-methods user-story \cite{WauteletHKK17} (column 3).

The first concern outlines CS-1: Given a significant event, data about the event, such as the number of people and hazardous places (i.e., hotspots), should be collected.

The second concern defines CS-2 and CS-5 about hotspot identification. 
CS-5 eases a security enforcer to identify hotspots employing predictive telemetry and visualizing maps, while CS-2 uses the generated telemetry to guide agents to the hotspots.

Additionally, two more scenarios, namely CS-3 and CS-7, were derived from the third concern. 
In particular, CS-3 aims at providing more insights about hotspots and anomalous behaviours from remote long-range videos in smartglasses and smartphones. Hence, security enforcers can keep track and monitor the hotspots previously mapped. 
CS-5 regards tracking anomalous behaviour in remote video streaming.

The fourth concern (CS-8) regards the pick-pocketing issue in social events. 
CS-8 implements the pick-pocketing scenario by producing notifications of anomalous trajectories in nearby sensitive areas, and a security agent can responsively reach the spot.

Finally, the last concern was mapped with two case scenarios.
On the one hand, CS-4 provides a remote video feed, helping the Paaspop organization keep track of critical case scenarios (e.g., moshpit, pick-pocketing).
On the other hand, CS-6 helps to manage particular crowd behaviour gathered with the CS-4.

\section{Research Design} \label{sec:research_design}

\begin{figure}[t]
    \centering
    \includegraphics[width=.7\linewidth]{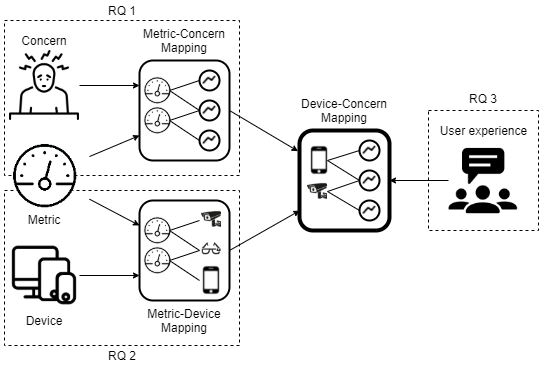}
    \caption{Pipeline of the device-concern mapping process.}
    \label{fig:paaspop_pipeline}
\end{figure}

\subsection{The VISOR Experience Report: Research Problem and Questions}
Our main goal in the scope of the VISOR field study was to create a framework capable of mapping stakeholders' concerns using IoT devices to monitor CPS and the spaces around them. Consequently, the study reported in this paper aims at providing architecture and infrastructure with a variety of devices and a data pipeline featuring the entire monitoring of the Paaspop event and processes around it as part of the cyber-physical space the event entails. The experience report featured in this manuscript started with looking at the concerns and major issues reported by the relevant stakeholders (e.g., hotspot identification, anomalous behaviours in a crowd) from their experience with previous editions. For example, on one side, a large field of view camera (e.g., drones, CCTV) is in place to recognize a pick-pocketing action analyzing people's anomalous trajectory behaviour. On the other side, a more dynamic device (e.g., smartphone, smart-glass) eases the recognition of pick-pocketing actions looking at the person's movement in a crowd. Subsequently, we aimed at creating a service platform, namely a software infrastructure and data pipeline after that IoT devices could interact with each other \cite{BaresiRS13}, aiming to improve the accuracy of the targeted hazard scenarios' recognition (e.g., pick-pocketing, moshpit detection \cite{byun2019creating}, terrorist attack on the crowd).
Finally, in order to evaluate the platform architecture proposed in this study, we used non-functional metrics outlined by previous work in Chung et Al. \cite{chung2012non} (see a detailed description in \Cref{sec:exp_setup}). 

This section provides details over the last two parts of the summary above, namely, the preparation and evaluation of the proposed infrastructure and pipeline, starting from an elaboration of the involved devices and experimented architectural dimensions.

First, the devices chosen are described in \Cref{tab:device_spec}. We chose these devices for two reasons: 1) the stakeholders provided us with a drone and Paaspop CCTV, and 2) we need something that can handle the static and dynamic scenario, with both a view from the top and directly from the ground. On the one hand, a large field of view eases anomalous behaviour detection because it is more important to view a big part of the event concerning the quality of the scenario recorder. However, on the other hand, a device with a low field of view ensures a high image quality if used in a closed environment. In this way, it is possible to detect small actions, as pick-pocketing, difficult to detect with other devices with a higher field of view.

The type of devices mentioned earlier can be split into two main categories: (a) user-dependent devices and (b) independent devices. Except for CCTV, where the only human interaction is for the maintenance process, all other devices involve a human factor. Thus, for example, the drone needs a pilot to be driven, the smart-glasses need to be worn by humans (e.g., a LEAs) as part of his/her \emph{sentry-mode} run across the Paaspop ground, and the smartphone needs to be pointed by humans in the desired direction, possibly in continuity with an additional human at the LEAs command \& control centre.


In the Paaspop event, we built an IoT infrastructure based on devices shown in \Cref{tab:device_spec}: x3 smartphones, x3 smartglasses, x3 CCTV on 35 available in the Paaspop event provided by Paaspop festival organization (x2 for each tent plus x5 fish-eye cameras \cite{Gennery06} for shared areas and open grounds), and x1 drone provided to LEAs by us in the scope of the VISOR project. 

Through these devices, a data pipeline was created to interconnect them with the stakeholders' concerns and enact AI-based detection scenarios using ad-hoc software developed in the two weeks before the event and in the first two days of the same event, keeping the final day as a test-bed for the overall solution.


\begin{figure}[t]
    \centering
    \includegraphics[width=.9\linewidth]{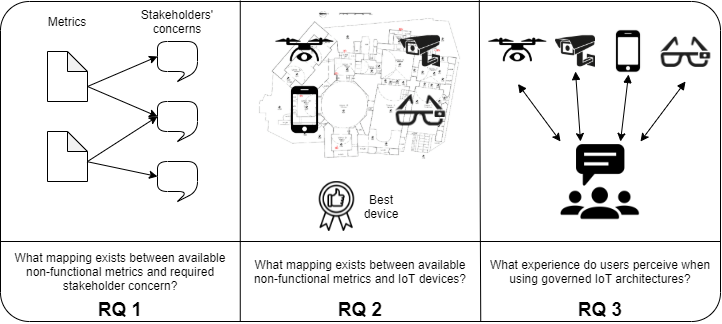}
    \caption{Overview of the research questions outlined for the Paaspop field study; RQs (bottom of the figure) are mapped to artefacts and data used for their evaluation (to part of the figure).}
    \label{fig:research_design_diagram}
\end{figure}
\begin{table}[b]
\centering
\footnotesize
\renewcommand{\arraystretch}{1.5}
\resizebox{0.85\linewidth}{!}{\begin{tabular}{lllllp{5cm}}
\toprule
\textbf{Class} & \textbf{Device} & \textbf{Quantity} & \textbf{Battery life} & \textbf{FOV} & \textbf{Other specs} \\
\midrule
Drone & DJI Matrice M200 V2 & 1 & 24 minutes & High & v \\
CCTV & - & 3 & Always on & Medium & N/A \\
Smartglass & VUZIX M400 & 3 & 45 minutes with the powerbank included & Low & x \\
Smartphone & Samsung Galaxy S10 & 3 & 8 hours in video recording & Low & CPU octacore 2.73GHz, RAM 8GB, internal memory 512GB, camera 12Mpx, front camera 10Mpx \\
\bottomrule
\end{tabular}
}
\caption{Devices' specifications.}
\label{tab:device_spec}
\end{table}

Hence, using our VISOR platform, every time a device recognizes illicit content, the VISOR server collects data from each device (e.g., videos, snapshots, GPS and camera footage from cellphones) and make an architectural screenshot of the actual system configuration, gathering metadata from each device (e.g., the position of each device, the autonomy). 
If illicit content is detected, the server sends a notification to all the smartphones and smart-glasses used by Paaspop's VISOR agents still active and walking in sentry mode on the ground. 

Considering the motivation raised and in the scope of the aforementioned experimentation, we defined the following main research question:

\begin{center}
    \textit{\textbf{Main RQ:} How can the most appropriate IoT architecture to secure cyber-physical spaces based on experienced stakeholders' concerns be designed?}    
\end{center}

To answer the main research question, we outlined three sub research questions:
\begin{itemize}
    \item{\textbf{RQ$1$}:} What mapping exists between available non-functional metrics and required stakeholder concern?
    
    
    \item{\textbf{RQ$2$:} What mapping exists between available non-functional metrics and IoT devices?}
    
    \item{\textbf{RQ$3$:} What experience do users perceive when using governed IoT architectures?}
\end{itemize}

\Cref{fig:paaspop_pipeline} provides a view of the pipeline process used to answer the main research question. It shows how to integrate the three RQs in order to find the best device-concern configuration. The idea is to cross-reference data gathered from RQ$1$ and RQ$2$ to interconnect concerns and devices and use the user feedback in the RQ$3$ to reinforce the mapping.
An overview of the three research questions is presented in \Cref{fig:research_design_diagram}. The first two RQs are strictly connected. If a stakeholder wants to know the best configuration to solve a concern (and, therefore, a set of case scenarios), the RQ$1$ creates a mapping between concerns and metrics, providing the right metric for a specific concern. Instead, the RQ$2$ provides a mapping between the metric and a set of possible devices useful with that metric. From the knowledge retrieved by these two RQs, the stakeholder knows the best devices for that metric. For example, if there is a mapping between the CS-4 (anomalous trajectory nearby sensitive area) and a metric, and the RQ$2$ suggests the usage of smartglasses for that metric, it means that smartglasses are the best device to solve that case scenario.
Regarding RQ$3$, we aim to provide suggestions to the stakeholders, gathering information from agents directly involved in the experimentation. For example, if the smart-glass eases the pick-pocketing recognition better than a smartphone, but agents find difficulties using it, choosing a smartphone might be a wiser choice.

\subsection{Experimental Setup and Methods}\label{sec:exp_setup}

Considering the RQs raised in \Cref{sec:research_design}, and considering that our PaasPop experimentation through VISOR involves a considerable amount of private or privacy-sensitive information, for each of said RQs, we were forced to adopt a specific and data-safe research strategy.

First, starting from the metrics outlined by Chung et Al. \cite{chung2012non}, we selected a limited set of non-functional requirements in concert with rankings operated by our reference stakeholders in the scope of the previously mentioned focus groups. Emerged requirements are reported in the following, tailoring a definition from the ISO/IEC 9126 software quality standard:

\begin{enumerate}
 
   \item \textbf{Usability}: is measured as the ease to use the technologies involved in the specific source. Examples of technologies can be language programming, software that requires specific skills, and lacking communication among devices;

   \item \textbf{Reliability}: measures how technology is fault tolerance from a software perspective. Hence, which technology provides the suitable characteristics to prevent disruptions arising from a single point of failure, ensuring the high availability and business continuity of mission-critical applications or systems;

   \item  \textbf{Adaptability}: is the ability of a technology or a system to adapt itself efficiently and quickly given the circumstances, from a hardware perspective;

   \item \textbf{Performance}: in this context, as a performance requirement, we are using the concept of autonomy. The definition of \textit{``autonomy''} can be summarized as the time that technology can keep the monitoring activity without interruptions. In this context, we analyze the technology considering the possibility to have an extra power supply like power banks or secondary batteries to extend battery life;
\end{enumerate}

It should be noted, however, that, although we have concentrated on the following list based on discussions/interviews with stakeholders (see \Cref{tab:stakeholders}), the technical availability of atomic measurements that the IoT devices had available forced us to concentrate on an even more limited set of non-functional requirements, relating primarily to energy-consumption, performance and other data available from the devices themselves.



\begin{figure}[t]
    \centering
    \includegraphics[width=\linewidth]{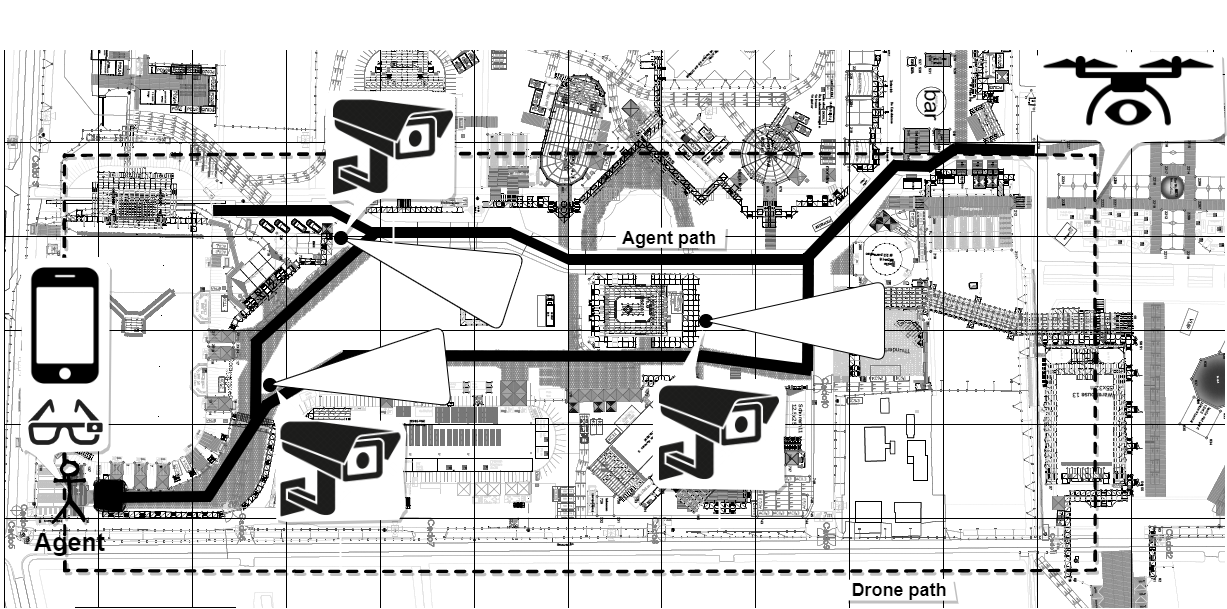}
    \caption{The dashed trajectory line is the path monitored by the drone, the solid one has been followed by it security staff, wearing smartglass and smartphone, the circle, instead, are the cctv camera.} 
    \label{fig:roadmap_task}
\end{figure}

\begin{figure}[t]
    \centering
    \includegraphics[width=.8\linewidth]{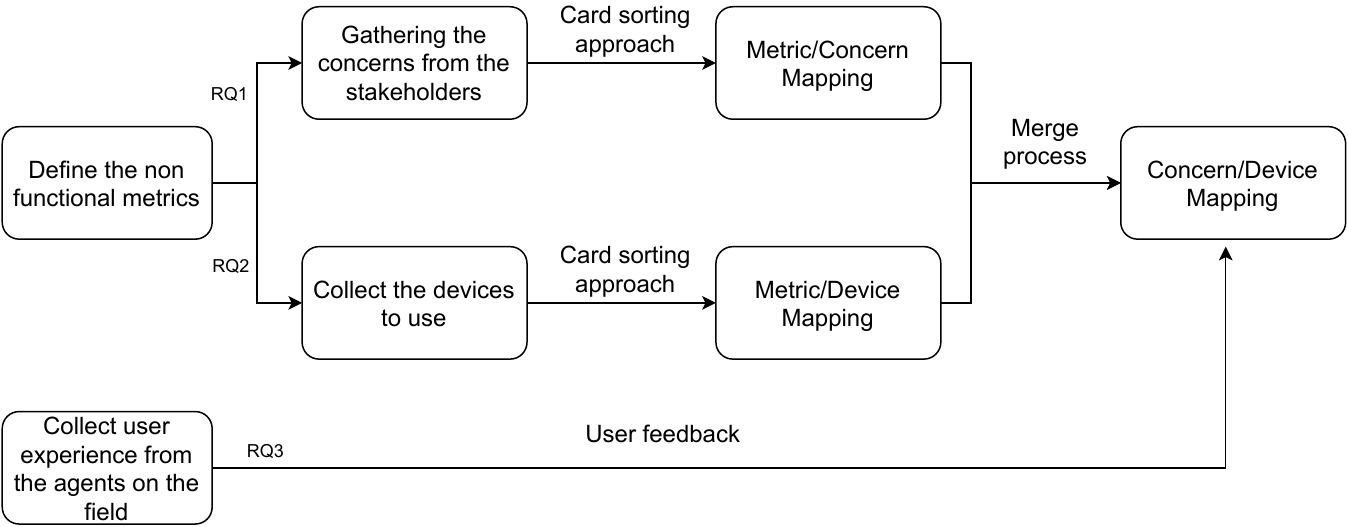}
    \caption{Flowchart of the device-concern mapping process, from the metrics definition to the merge process of the outcome of the RQ$1$ and RQ$2$ with external user feedback (RQ$3$) to validate the results achieved.}
    \label{fig:flowchart}
\end{figure}

Subsequently, to obtain the mapping between concerns and metrics (RQ$1$), we scanned each concern available and the set of device-specific metrics that could be taken into account in the context of our field experimentation. More specifically, in this step of the process, we considered the metrics that fit best with the primary concerns that emerged in the CPS context \cite{shi2011survey}. Afterwards, a card-sorting approach \cite{lewis2010open} was followed to identify the mapping between the concerns and the metrics chosen. The card sorting approach was operated by two authors of this work independently and later compared using the well-known Krippendorf Alpha coefficient \cite{Krippendorf80} for observation agreement. The $\alpha$ score essentially measures a confidence interval score stemming from the agreement of values across two distinctly reported observations about the same event or phenomenon. In our case, the value was applied to measure the agreement between configuration details, analyses, and statistical results of our analysis. The value was calculated initially to be 0.83, hence $\alpha>.800$, a standard reference value for highly-confident observations.  Subsequently, the value was used to drive the agreement between the two analyses up towards total alignment.



Subsequently, to answer the RQ$2$, we gathered relevant usage statistics and other data during the entire Paaspop event and our experimentation. More specifically, we created a dataset where each parameter provides information useful for each metric. The dataset description is provided in \Cref{sec:dataset}. 
 
Finally, regarding RQ$3$, we focused on performing a user analysis methodology tailored from Albert et al. in \cite{albert2013measuring}. 
Albert et al. define 11 different metrics to measure the user experience and provide different profiles where these metrics can be used. Considering the technologies involved and the role of agents who uses them, we selected five metrics that fit better with our study:

\emph{1)} \textbf{Task success}: used to know if a task has been completed successfully;

\emph{2)} \textbf{Task time}: expressed as the time need to complete a specific task;

\emph{3)} \textbf{Efficiency}: measures the user feedback based on the battery consumption and the level of quality of the streaming video;

\emph{4)} \textbf{Learnability}: refers to the learning level of user respect to a specific device; 

\emph{5)} \textbf{Self Reported Metrics}: awareness and usefulness.

We built an experimentation testbed based on the sentry patrol data during the entire event monitoring. More specifically, a task was assigned to each agent, consisting of a path to follow (black line in \Cref{fig:roadmap_task}) to travel along such a path for 30 minutes with one or more wearable devices (smartphone and smartglasses). Each path can be travelled based on travel times, 1, 2, or 3 times (respectively in 30 minutes, 1 hour, and 1 hour and 30 minutes). 
More info about the tasks schedule can be found in the user experience Table (see \Cref{tab:user_exp}).
A task is considered performed successfully if the agent did not have any problems during its execution (e.g., full memory storage, data battery drained).
At the end of the experimentation, we interviewed each agent, asking them to answer every metric provided. 

\textcolor{\editcolor}{In summary, the workflow of the VISOR project along with all its components outlined previously, is recapped in \Cref{fig:flowchart}. The first step consists of defining non-functional metrics used for the mapping process between architecture concerns and physical (e.g., IoT) devices. In the next step, the critical set of concerns and devices are collected to answer, respectively, RQ$1$ and RQ$2$. After such gathering process, a card-sorting approach is applied to find a Metric-Concern mapping, to address RQ$1$, and a metric-device mapping to address RQ$2$. The next step provides the merging of the mapping found before to find a correlation between concerns and devices. In parallel, the experimentation involves the collection of user experience from the agents on the field regarding the usage of the wearable devices in order to validate the final mapping defined for the main RQ.}






\begin{table*}[h!]
\centering
\footnotesize
\renewcommand{\arraystretch}{1.5}
\resizebox{0.85\linewidth}{!}{
\begin{tabular}{ll}
\toprule
\textbf{Dataset parameter} & \textbf{Description} \\
\midrule
Scenario ID & Case scenario ID. \\ 
Device & Device implied in the experimentation.\\
Network in use & Network used for the experiment (e.g., WIFI, 4G, Bluetooth).\\
Timestamp & Timestamp of the experimentation.\\
Bandwidth & Bandwidth of the device(s), measured in the communication within the server or between each other.\\
Packet-loss & Number of packet loss during the communication with the internal server.\\
Latency & Delay between the device and the internal server.\\
Operator conditions & Quality level of the wireless connection (e.g. the WIFI or 4G signal power).\\
Object Telemetry & Hardware telemetries for each device (e.g. CPU usage, TEMP level).\\
Battery-Level & Level of the battery, for the entire experimentation.\\
Object trajectory & Level of deviance from the regular trajectory.\\

\bottomrule
\end{tabular}
}
\caption{Dataset description containing identification and quantitative parameters used to evaluate devices.}
\label{tab:dataset_params}
\end{table*}

\subsection{Dataset overview} \label{sec:dataset}


\begin{table}[h!]
\centering
\footnotesize
\renewcommand{\arraystretch}{1.5}
\resizebox{0.85\linewidth}{!}{
\begin{tabular}{lcccccccc}
\toprule
\textbf{Metric} & \textbf{Network in use} & \textbf{Bandwidth} & \textbf{Packet-loss} & \textbf{Latency} & \textbf{Operator conditions} & \textbf{Object telemetry} & \textbf{Battery level} & \textbf{Object trajectory} \\
\toprule
\textbf{Usability} & \cmark & \xmark & \xmark & \xmark & \xmark & \xmark & \xmark & \xmark \\
\textbf{Performance} & \xmark & \xmark & \xmark & \xmark & \xmark & \xmark & \cmark & \xmark \\
\textbf{Adaptability} & \xmark & \xmark & \xmark & \xmark & \xmark & \xmark & \xmark & \cmark \\
\textbf{Reliability} & \xmark & \cmark & \cmark & \cmark & \cmark & \cmark & \xmark & \xmark \\
\bottomrule
\end{tabular}
}
\caption{Mapping between the dataset parameters and the non functional metric analyzed.}
\label{tab:dataset_metric}
\end{table}

Considering the non-functional metrics described in \Cref{sec:exp_setup}, we created a dataset containing parameters useful for the technology categorization, evaluating the fitness of a device for each metric defined.

We provide a list of the parameters with a description in \Cref{tab:dataset_params}.
The parameters described can be grouped into two categories:

\begin{enumerate}
    \item \textbf{Identification parameters:} Scenario ID, device, and timestamp are used to identify a specific scenario configuration (e.g., tracking a critical scenario's evolution (on CS-4) with a smart-glass in a specific time). It means that each row in the dataset refers to a specific case scenario and a device used in it.
    \item \textbf{Quantitative parameters:} The other parameters in \Cref{tab:dataset_params} are used to gather the information to map a device with a specific metric. 
    For example, to detect an anomalous behaviour (as a thief running away), the device tracks the trajectory of every person. The difference between the trajectories recorder and the regular one is saved in the "Object trajectory" parameter.
\end{enumerate}

Each quantitative parameter reflects one measurable quantity emitted by the targeted devices. Thus, if there is a mapping between the battery level parameter and the performance metric, it is possible to evaluate the device used in the case scenario. 



The mapping between the quantitative parameter discussed and the metrics explained in \Cref{sec:exp_setup} is shown in \Cref{tab:dataset_metric}.

As we can see in the table, a parameter can only connect with a metric. Differently, a metric can have connections with more parameters.
The choice of the parameter outlined was taken after discussing the limitation of devices used and the need to find a mapping on at least one parameter for each metric. For example, we can only use the battery life as an estimation of the performance of a device because the Android OS limitations, after 8.0, do not allow to retrieve other helpful information, as CPU usage or RAM usage \cite{android-issue}.
\section{Results} \label{sec:results}
\begin{figure}[h!]
    \centering
    \includegraphics[width=.75\linewidth]{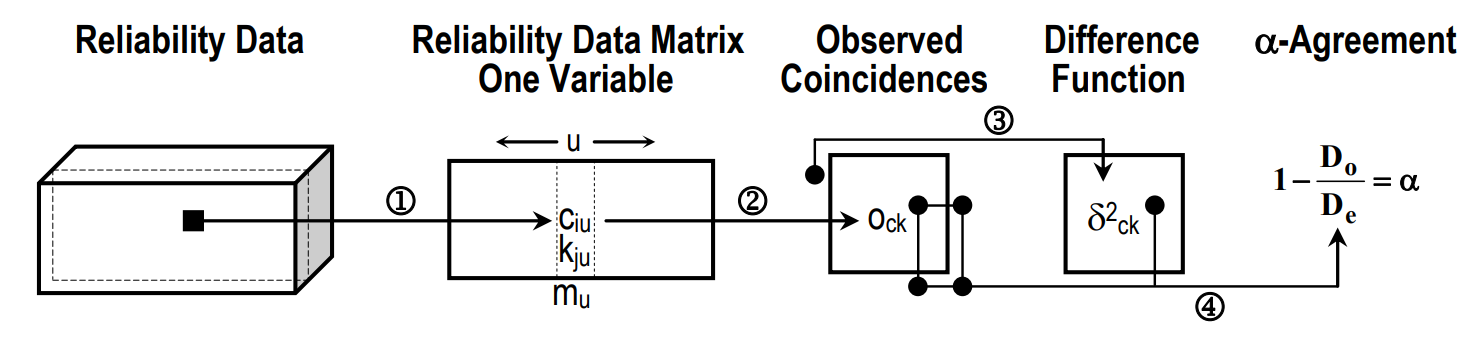}
    \caption{The process consists in four steps that, starting from the reliability data, estimates the $\alpha$-coefficient.
    }
    \label{fig:krippendorff-pipeline}
\end{figure}

This section shows the results for the research questions presented in \Cref{sec:introduction}. The three research questions led us to establish the metric-concern/device mapping relation to address the main research goal.

For the first two RQs, we enacted a card sorting approach to define a mapping between metrics and concerns (RQ$1$) and the relationship between metrics and devices (RQ$2$). 
First, we opened a joint discussion on the list of concerns, devices, and metrics outlined in the previous sections to establish a rational connection that may impact the CPS. 
Afterwards, we compared the result using Krippendorff's alpha~\cite{krippendorff2011computing} to measure the level of agreement among several observers. 
Given the data extracted during the discussion, the $\alpha$-coefficient can be computed using the process shown in \Cref{fig:krippendorff-pipeline}, defined in Krippendorff's work \cite{krippendorff2011computing}.

The process in \Cref{fig:krippendorff-pipeline} can be defined with four different types of data: 1) binary data, two observers, no missing data, 2) nominal data, two observers, no missing data, 3) nominal data, multiple observers (>2), missing data, 4) all metrics, any number of observers, missing data.

Instead, the RQ$3$ uses five metrics (Task status, task time, efficiency, learnability, and self-reported metric) to evaluate user feedback of our agents.






\subsection{RQ1: What mapping exists between available non-functional metrics and required stakeholder concern?}






\noindent\begin{minipage}{\linewidth}
\centering
\footnotesize
\renewcommand{\arraystretch}{1.5}
\begin{tabular}{lccccc|ccccc|ccccc|ccccc}
\hline
\multicolumn{1}{l}{} & \multicolumn{5}{c}{\textbf{Usability (M$1$)}} & \multicolumn{5}{c}{\textbf{Performance (M$2$)}} & \multicolumn{5}{c}{\textbf{Adaptability (M$3$)}} & \multicolumn{5}{c}{\textbf{Reliability (M$4$)}} \\
\toprule
& \textbf{C$1$} & \textbf{C$2$} & \textbf{C$3$} & \textbf{C$4$} & \textbf{C$5$} & \textbf{C$1$} & \textbf{C$2$} & \textbf{C$3$} & \textbf{C$4$} & \textbf{C$5$} & \textbf{C$1$} & \textbf{C$2$} & \textbf{C$3$} & \textbf{C$4$} & \textbf{C$5$} & \textbf{C$1$} & \textbf{C$2$} & \textbf{C$3$} & \textbf{C$4$} & \textbf{C$5$} \\
\textbf{A$1$} & 0 & 0 & 1 & 0 & 0 & 1 & 0 & 0 & 1 & 1 & 0 & 0 & 0 & 1 & 1 & 1 & 1 & 0 & 1 & 1 \\
\textbf{A$2$} & 0 & 0 & 1 & 0 & 0 & 1 & 1 & 0 & 1 & 1 & 0 & 0 & 0 & 1 & 1 & 1 & 0 & 0 & 1 & 1 \\
\bottomrule
\end{tabular}
\captionof{table}{Reliability data matrix obtained from the agreement data gathered by the authors. A$1$ and A$2$ represent, respectively, the first and the second author. C$1$-$5$ represent the concerns for each metric analyzed (M$1$-$4$).}
\label{tab:rel_data_matrix}
\end{minipage}

\begin{figure}[h]
    \centering
    \includegraphics[width=.5\linewidth]{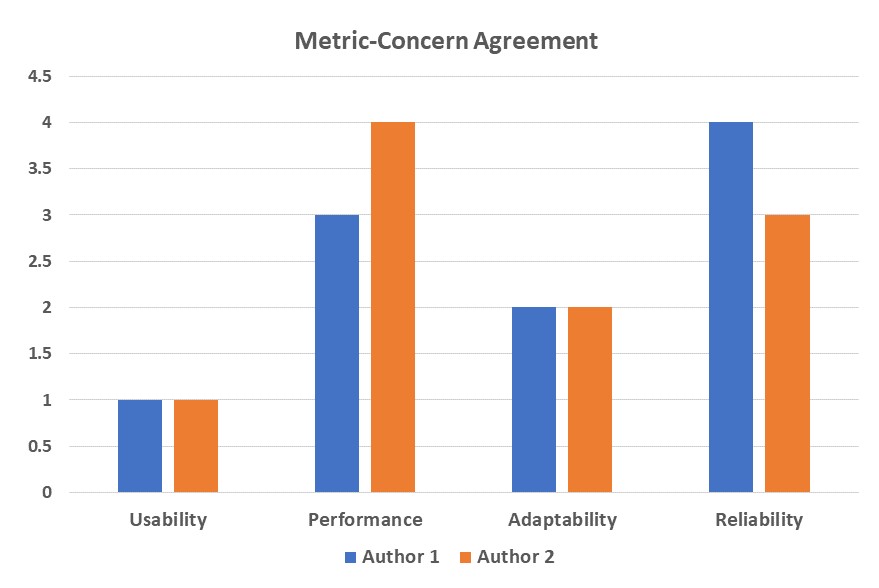}
    \caption{\textcolor{\editcolor}{Histogram of the agreement data collected by raters in \Cref{tab:rel_data_matrix}. For each metric, they have been counted the number of positive values in the reliability data matrix of the authors.}}
    
    \label{fig:histogram_rq1}
\end{figure}

Considering the data supported by the process in \Cref{fig:krippendorff-pipeline}, our study lies with the first scenario. The data collected are binary because each concern may have a relationship with a specific metric. Therefore, the observer considers the first two authors of this work, and there are no missing data.
\textcolor{\editcolor}{The histogram shown in \Cref{fig:histogram_rq1} depicts the number of positive agreements of the raters for each metric. In this context, regarding usability and adaptability, both raters have the same level of agreement.}

The first step of the process, starting from the agreement data collected by the authors (A$1$ and A$2$), constructs a reliability data matrix shown in \Cref{tab:rel_data_matrix}. For each metric-concern pair, we record the agreement in a binary form. $0$ if there is a disagreement on the pair, $1$ if there is an agreement. The two observer matrix contains $N=20$ metric-concern pairs, for a total of $40$ values.

\begin{table}[ht!]
\centering
\footnotesize
\renewcommand{\arraystretch}{1.5}
\resizebox{.5\linewidth}{!}{
\begin{tabular}{ccccc}
                       & 0  & 1                       &    &  \\ \cline{2-3}
\multicolumn{1}{l|}{0} & \textit{o}$\mathit{_{00}}$ & \multicolumn{1}{l|}{\textit{o}$\mathit{_{01}}$}  & \textit{n}$\mathit{_0}$ &  \\
\multicolumn{1}{l|}{1} & \textit{o}$\mathit{_{10}}$  & \multicolumn{1}{l|}{\textit{o}$\mathit{_{11}}$} & \textit{n}$\mathit{_1}$ &  \\ \cline{2-4}
                       & \textit{n}$\mathit{_0}$ & \multicolumn{1}{l|}{\textit{n}$\mathit{_1}$} & \textit{n} $= 2$\textit{N} & 
\end{tabular}

\begin{tabular}{ccccc}
                       & 0  & 1                       &    &  \\ \cline{2-3}
\multicolumn{1}{l|}{0} & 18 & \multicolumn{1}{l|}{2}  & 20 &  \\
\multicolumn{1}{l|}{1} & 2  & \multicolumn{1}{l|}{18} & 20 &  \\ \cline{2-4}
                       & 20 & \multicolumn{1}{l|}{20} & 40 & 
\end{tabular}
}
\caption{The confidence matrix tabulates the 20 pairs recorded in the reliability data matrix in a 2 by 2 matrix.}
\label{tab:coincidence_matrix}
\end{table}

In the second step, from the reliability data matrix, we created the coincidences matrix in \Cref{tab:coincidence_matrix}. Accordingly with the table, \textit{o}$\mathit{_{00}}$ represents the 9 pairs 0-0, 
\textit{o}$\mathit{_{01}}$ and \textit{o}$\mathit{_{10}}$ represent the disagreement pairs (0-1 and 1-0) in the reliability matrix and \textit{o}$\mathit{_{11}}$ represents the 9 pairs 1-1, \textit{n}$\mathit{_0}$ and \textit{n}$\mathit{_1}$ are the sum of, respectively, 0 and 1 in the reliability matrix and n=2N is the sum of the agreement values expressed in \Cref{tab:rel_data_matrix}.

The difference function is calculated in step 3. As explained by Krippendorff, it was skipped because it is not necessary to use a different function for the data that we have.

In the final step, $\alpha$-coefficient has been calculated as follows:


\begin{equation}
binary \; \alpha = {1 - (n-1)} \cdot {o_{01}\over{n_0 \cdot n_1}}
\end{equation}



An $\alpha$ value of 1 means that there is a complete agreement between the observers. Conversely, a value of 0 indicates the absence of reliability in the data gathered. Considering that an acceptable level of agreement is when $\alpha > 0.800$ \cite{krippendorff2004reliability}, we estimated
\textcolor{\editcolor}{, based on the data in \Cref{tab:coincidence_matrix},} an $\alpha$ value of $0.805$.
\textcolor{\editcolor}{This value is in line with the outcome of the histogram in \Cref{fig:histogram_rq1}, considering the closeness of the raters' agreement.}

\begin{table*}[ht!]
\centering
\footnotesize
\renewcommand{\arraystretch}{1.5}
\begin{tabular}{lrrrrrrr}
\toprule
   & \textbf{C$1$} & \textbf{C$2$} & \textbf{C$3$} &  \textbf{C$4$} & \textbf{C$5$} \\
\toprule
    \textbf{Usability} & \xmark & \xmark & \cmark & \xmark & \xmark \\
    \textbf{Performance} & \cmark & \xmark & \xmark & \cmark & \cmark \\
    \textbf{Adaptability} & \xmark & \xmark & \xmark & \cmark & \cmark \\
    \textbf{Reliability} & \cmark & \xmark & \xmark & \cmark & \cmark \\
\bottomrule
\end{tabular}

\caption{\textcolor{\editcolor}{Concern/Metric score matrix. The results of this table are derived from the reliability data matrix. Checkmarks indicated an agreement has been reached between authors.}}
\label{tab:rq1_score_matrix}
\end{table*}

\textcolor{\editcolor}{At the end of the process, the metric-concern pairs chosen based on the results in \Cref{tab:rel_data_matrix} outlined in \Cref{tab:rq1_score_matrix}, was mapped by the following:}


\begin{itemize}
    \item \textbf{Usability:} We found a mapping between this metric and concern 3 (case scenarios 3 and 7). We evaluated the usability in the communication process between smartphones and smartglasses and the internal server. When the CCTV sends data to the server, the data are analyzed to find some hazard areas, possible hotspots, and anomalous trajectories. If the server recognizes these scenarios, it sends a notification to our agents' smartphones and smartglasses. 
    \item \textbf{Adaptability:} The best concerns that fit with this metric are concerns 4 and 5. Both of them manage special crowd behaviour, critical scenario, and anomalous trajectory. Accordingly, a device with a high level of adaptability can better handle these critical scenarios.
    \item \textbf{Reliability:} This metric is linked with concerns 1, 4, and 5 because the case scenario (case scenarios 1, 4, 6 and 8) provides our devices' direct usage. In Paaspop's context, we analyzed how our device was reliable during the monitoring phase (e.g., issues in the Bluetooth communication between smartphone and smart-glass, full memory storage).
    \item \textbf{Performance:} This metric, such as the reliability metric, has a mapping with concerns 1, 4, and 5. Being the only concern that implies a direct usage of our devices, we can evaluate their autonomy during the experimentation. 
\end{itemize}

\subsection{RQ2: What mapping exists between available non-functional metrics and IoT devices?}


\noindent\begin{minipage}{\linewidth}
\centering
\footnotesize
\renewcommand{\arraystretch}{1.5}
\begin{tabular}{lcccc|cccc|cccc|cccc}
\toprule
\multicolumn{1}{l}{} & \multicolumn{4}{c}{\textbf{Usability (M$1$)}} & \multicolumn{4}{c}{\textbf{Performance (M$2$)}} & \multicolumn{4}{c}{\textbf{Adaptability (M$3$)}} & \multicolumn{4}{c}{\textbf{Reliability (M$4$)}} \\
\toprule
& \textbf{D$1$} & \textbf{D$2$} & \textbf{D$3$} & \textbf{D$4$} & \textbf{D$1$} & \textbf{D$2$} & \textbf{D$3$} & \textbf{D$4$} & \textbf{D$1$} & \textbf{D$2$} & \textbf{D$3$} & \textbf{D$4$} & \textbf{D$1$} & \textbf{D$2$} & \textbf{D$3$} & \textbf{D$4$} \\
\textbf{A}$1$ & 5 & 5 & 3 & 1 & 4 & 3 & 5 & 1 & 5 & 5 & 2 & 4 & 2 & 2 & 5 & 5 \\
\textbf{A}$2$ & 5 & 5 & 2 & 1 & 4 & 3 & 5 & 1 & 5 & 5 & 2 & 2 & 2 & 3 & 5 & 5 \\
\bottomrule
\end{tabular}
\captionof{table}{Reliability data matrix obtained from the agreement data gathered by the authors. A$1$ and A$2$ represent, respectively, the first and the second author. C$1$-$5$ represent the concerns for each metric analyzed (M$1$-$4$).}
\label{tab:rel_data_matrix_rq2}
\end{minipage}

\begin{figure}[h]
    \centering
    \includegraphics[width=.5\linewidth]{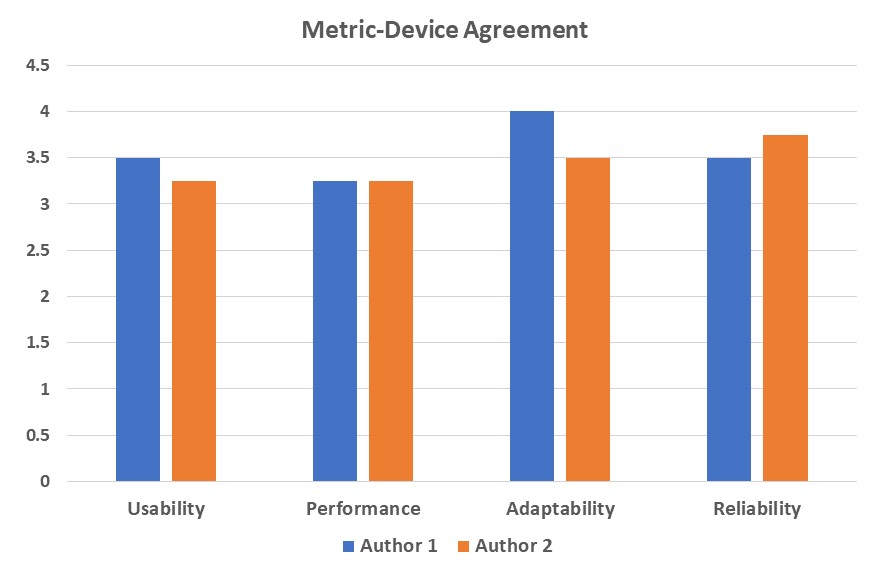}
    \caption{\textcolor{\editcolor}{Histogram of the agreement data collected from raters in \Cref{tab:rel_data_matrix_rq2}. For each metric, the average of the values in the reliability data matrix of the raters was counted and reported.}}
    
    \label{fig:histogram_rq2}
\end{figure}


Differently from the RQ$1$, the agreement data collected are nominal. They are based on the following classification:

\begin{enumerate}
    \item \textbf{Very Bad:} a device is useless for the targeted purpose (or raises problematic issues), and its usage could harm the experiment's final goal.
    
    \item \textbf{Bad:} the device cannot provide enough useful data, or it is problematic to use it (or access to it).
    
    \item \textbf{Neutral:} the benefits and the issues are the same. 
    
    \item \textbf{Good:} the device is easy to use, has no specific issues, and can extract enough data without specific problems.
    
    \item \textbf{Very Good:} the device can perfectly perform its task. There are no issues (bureaucratic and technological), and it is easy to use or access.
\end{enumerate}

Hence, our study follows the second data scenario, where nominal data with two authors and no missing data are used for the agreement measurement. 
\textcolor{\editcolor}{The histogram shown in \Cref{fig:histogram_rq2} depicts the average of the raters' agreements for each metric. In this context, the level of agreement of each raters is close to the other.}

Firstly, after having gathered the agreement data from the first two authors of this work (A$1$ and A$2$), we built the reliability data matrix shown in \Cref{tab:rel_data_matrix_rq2}. We saved the agreement in a numerical form for each metric-device pair, following the classification shown ahead. The matrix contains $N=16$ metric-concern pairs, for a total of $32$ values.

\begin{table}[h!]
\centering
\footnotesize
\renewcommand{\arraystretch}{1.5}
\resizebox{.6\linewidth}{!}{
    \begin{tabular}{ccccccc}
                           & 1 & . & k & . & .                       &    \\ \cline{2-6}
    \multicolumn{1}{c|}{1} & \textit{o$\mathit{_{11}}$} & . & \textit{o$\mathit{_{1k}}$} & . & \multicolumn{1}{c|}{.}   & \textit{n}$\mathit{_1}$  \\
    \multicolumn{1}{c|}{.} & . & . & . & . & \multicolumn{1}{c|}{.}   & .  \\
    \multicolumn{1}{c|}{.} & . & . & . & . & \multicolumn{1}{c|}{.}   & .  \\
    \multicolumn{1}{c|}{c} & \textit{o$\mathit{_{c1}}$} & . & \textit{o$\mathit{_{ck}}$} & . & \multicolumn{1}{c|}{.}   & \textit{n}$\mathit{_c} = \sum_c$ \textit{o}$_{ck}$  \\
    \multicolumn{1}{c|}{.} & . & . & . & . & \multicolumn{1}{c|}{.} & . \\ \cline{2-7} 
                           & \textit{n$\mathit{_1}$} & . & \textit{n$\mathit{_k}$} & . & \multicolumn{1}{c|}{.} & $\textit{n} = \sum_c \sum_k$ \textit{o}$_{ck}$
    \end{tabular}
    
    \begin{tabular}{ccccccc}
                           & 1 & 2 & 3 & 4 & 5                       &    \\ \cline{2-6}
    \multicolumn{1}{c|}{1} & 4 & . & . & . & \multicolumn{1}{c|}{.}   & 4  \\
    \multicolumn{1}{c|}{2} & . & 4 & 2 & 1 & \multicolumn{1}{c|}{.}   & 7  \\
    \multicolumn{1}{c|}{3} & . & 2 & 2 & . & \multicolumn{1}{c|}{.}   & 4  \\
    \multicolumn{1}{c|}{4} & . & 1 & . & 2 & \multicolumn{1}{c|}{.}   & 3  \\
    \multicolumn{1}{c|}{5} & . & . & . & . & \multicolumn{1}{c|}{14} & 14 \\ \cline{2-7} 
                           & 4 & 7 & 4 & 3 & \multicolumn{1}{c|}{14} & 32
    \end{tabular}
}

\caption{The confidence matrix tabulates the 16 pairs recorded in the reliability data matrix in a 5 by 5 matrix.}
\label{tab:coincidence_matrix-rq2}
\end{table}

Secondly, based on the reliability data matrix in \Cref{tab:rel_data_matrix_rq2}, we outlined the coincidences matrix in \Cref{tab:coincidence_matrix-rq2}. It represents a five-by-five matrix because there are five types of classification. Each \textit{o} value represents the number of pairs of the reliability matrix and is entered twice. For example, $\textit{o}\mathit{_{12}}$ indicates the 1-2 and 2-1 pairs. Conversely, n$_c$ value represents the sum of all o$_{c}$ values, while n value represents the sum of all n values.

Lastly, we estimated the $\alpha$ coefficient by the follows:


\begin{equation}
nominal \; \alpha = {{(n-1) \sum_c o_{cc} - \sum_c n_c(n_c-1)}\over{n(n-1) - \sum_c n_c(n_c-1)}}
\end{equation}

After evaluating the goodness of the agreement 
\textcolor{\editcolor}{with an $\alpha$ coefficient of $0.748$,}
we outlined the results in \Cref{tab:score_matrix} applying the arithmetic mean on the reliability matrix in \Cref{tab:rel_data_matrix_rq2}. If a value is not an integer, it is rounded down.
\textcolor{\editcolor}{The $\alpha$ coefficient value is justified by the outcome of the histogram in \Cref{fig:histogram_rq2}, considering the closeness of the average level registered between the raters.}

\begin{table*}[ht!]
\centering
\footnotesize
\renewcommand{\arraystretch}{1.5}
\begin{tabular}{lrrrrrr}
\toprule
   & \textbf{Smartphone} & \textbf{Smartglass} & \textbf{CCTV} &  \textbf{Drone} \\
\toprule
                  \textbf{Usability} & 5 & 5 & 2 & 1 \\
                \textbf{Performance} & 4 & 3 & 5 & 1 \\
               \textbf{Adaptability} & 5 & 5 & 2 & 3 \\
                \textbf{Reliability} & 2 & 2 & 5 & 5 \\
\midrule
Average & 4 & 3.75 & 3.5 & 2.5 \\
\bottomrule
\end{tabular}

\caption{Matrix Technology/Metric score. 1 = Very Bad, 2 = Bad, 3 = Neutral, 4 = Good, 5 = Very Good.}
\label{tab:score_matrix}
\end{table*}


Considering the data in \Cref{tab:score_matrix}, we found the following relations:

\begin{itemize}
    \item \textbf{Usability}: in this study, we demonstrated that the most usable sources are smartphones and the smart-glasses. Since they mount a Linux-based Operating System, namely Android, it can enhance the OS environment with new features implementing applications written in Java. Moreover, Android OS allows building a federated learning framework to train an algorithm across multiple decentralized edge devices using the data gathered in each of them~\cite{yang2019federated}. Lastly, it is possible to build in the smartphone and smartglasses a real-time monitoring tool. Hence, the security guys can operate using the wearable devices live from the field without the intervention of the IT department that can continue monitoring and coordinating all the security guys' actions.
    Conversely, drones and CCTV present some technological issues. The security IT department that manages all the cameras can not provide direct access to the stream data without API and personal credentials for the CCTV. We retrieved stream data using the Paaspop organization's live stream platform, bringing additional costs to develop other tools for gathering such data. It has not been possible to use real-time data concerning the Drone because it has provided no solutions. Consequently, making impossible the real-time monitoring directly from the drone device.
    
    \item \textbf{Performance (autonomy)}: CCTV is the only data source from our list independent from battery power. Hence, if there are no power supply issues in the leading supply network (e.g., blackout or any tampering attempts), the CCTV can be considered the best technology to put in place concerning the autonomy perspective.
    Right after the CCTV, under the autonomy perspective, we have smartphones. They can keep recording for at least half a day~\cite{batterytest} and use ``normal'' (480p and 720p) video quality options.
    Conversely, smartglasses can provide considerably little time autonomy due to the small battery capacity. 
    As the latest technology under the autonomy perspective, we have the Drone. It is well known that drones need a balance between the weight of the battery and its capacity. The trade-off usually results in lightweight batteries providing small capacity, hence less autonomy life. The battery life is around ~20 minutes, making this type of technology the worst list. Hence, we found that a considerable limitation is a necessity of changing batteries every 20 minutes and producing an interruption of data flow. 
    
    \item \textbf{Adaptability}: among all the technologies we tested, we demonstrated that the devices with a higher level of Adaptability are smartphones and smartglasses. The portability of these devices makes them agile and easy to move in crowd scenes among people and gives the operator the possibility to move fast from one side to another.
    Drones need to be piloted by an expert; however, they can quickly and rapidly move on the target from the top. Therefore, while considering the Field of View (only from the top) and the critical battery limitations, we can not consider this type of technology reliable enough to guarantee good public space protection over an extended time.
    Lastly, we have CCTV. Our use study has been proved to be the most useless technology, given its static nature. Indeed, most of the time, the illicit crowd behaviours are not taken in front of the camera. 
    
    \item \textbf{Reliability}: during the festival, we have been able to notice that smartglasses and smartphones are the two technologies inclined to discontinuity. WiFi issues and full memory storage caused a loss of data. Moreover, the technologies suffered from crashes due to the high temperature of the devices. 
    Conversely, the experiments showed that both CCTV and drones proved an excellent level of robustness. For CCTV, all the architecture behind it has already been validated in the past. Regarding the Drone, instead, the battery limitation ensures that the memory storage cannot be full.
\end{itemize}

\subsection{RQ3: What experience do users perceive when using governed IoT architectures?}


\begin{table}[h!]
\centering
\footnotesize
\renewcommand{\arraystretch}{1.5}
\begin{tabular}{lllll}
\toprule
 \textbf{Agent}  & \textbf{Task status} & \textbf{Task time} & \textbf{Efficiency} & \textbf{Learnability}\\
\toprule
Agent 1 - Smartphone & Succeeded & 1h30m & Good & Easy \\
Agent 1 - Smartglass & Succeeded & 30m & Good & Device difficult to setup \\
Agent 2 - Smartphone & Succeeded & 1h & Good & Easy \\
Agent 2 - Smartglass & Failed & 1h & At limit, without external powerbank & Device difficult to setup \\
Agent 3 - Smartphone & Succeeded & 1h & Good & Easy \\
Agent 3 - Smartglass & Succeeded & 1h & At limit, without external powerbank & Device difficult to setup \\
\bottomrule
\end{tabular}

\caption{User experience of three agents during the experimentation.}
\label{tab:user_exp}
\end{table}

\Cref{tab:user_exp} shows the information gained by the user feedback for each metric shown in \Cref{sec:exp_setup} (except for the "Self-reported metric" that is described in detail below).

\begin{enumerate}
    \item \textbf{Smartphone}
    
    \emph{Agent 1-2-3:} Except for the task performed, their feedback is the same. They have successfully performed the task assigned without any specific problem. No specific concerns have been raised about the battery drain, so they did not read extra power to complete the task. The monitoring software provided has been easy to set up and to use.
    
    \item \textbf{Smartglass}
    
    \emph{Agent 1:} With a 30 minute task, there was no specific issue related to the battery drain. Unfortunately, we found difficulties accessing the application and executing it because of the small display and general misunderstanding of the usage of the buttons on the smart glass. Nevertheless, the technology has successfully performed the task.
    
    \emph{Agent 2:} For a task long more than 30 minutes, there were battery drain issues, requiring an additional power bank to keep monitoring the scene. Like the previous one, we found the same problem with accessing the application and executing it because of the small display and general misunderstanding of the usage of the buttons on the smart glass. Unfortunately, the task has not been executed. In the middle of the roadmap, the application crashed and, under instruction, it was impossible to restart it.
    
    \emph{Agent 3:} Same concern of agent 1, except for the learnability.  In this case, we found no problem with the display size of the smart glass, but the same concerns have been shared about the setup and the device's access.    
\end{enumerate}



Regarding the ``Self-reported metric'', all of them agree on the same point: the glasses are helpful to receive insights at runtime because they can be worn, but, unfortunately, the display is too small, so they found it difficult stressful to look into it. Instead, the smartphone is valuable and reliable, but it is not practised, considering that you need to hold it in your hand, pointing at the crowd, all the time.

\subsection{Main Research Question: How can the most appropriate IoT architecture to secure cyber-physical spaces based on experienced stakeholders' concerns?}


\begin{table}[h!]
\footnotesize
\renewcommand{\arraystretch}{1.5}
    \begin{subtable}{.5\textwidth}
    \centering
    \begin{tabular}{ll}
        \toprule
        \textbf{Metric} & \textbf{Concerns} \\
        \toprule
        Usability & Concern 2 and 3 \\
        Performance & Concern 1, 4 and 5 \\
        Adaptability & Concern 4 and 5 \\
        Reliability & Concern 1, 4 and 5 \\
        \bottomrule
    \end{tabular}
    \caption{}
    \label{subtab:mapping-rq1}
    \end{subtable}%
    \begin{subtable}{.5\textwidth}
    \centering
    \begin{tabular}{ll}
        \toprule
        \textbf{Metric} & \textbf{Best Device} \\
        \toprule
        Usability & Smartphone \& Smart-Glass \\
        Performance & CCTV \\
        Adaptability & Smartphone \& Smart-Glass \\
        Reliability & CCTV \& Drone \\
        \bottomrule
    \end{tabular}
    \caption{}
    \label{subtab:mapping-rq2}
    \end{subtable}
    \caption{Relation between metrics and concerns resulting from the RQ$1$ (a) and metrics and devices resulting from the RQ$2$ (b).}
\end{table}

\begin{figure*}[htp]
  \centering
  \includegraphics[width=.5\linewidth]{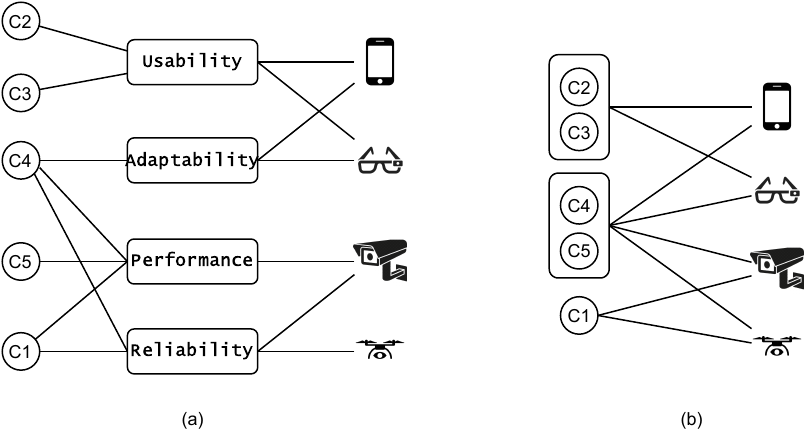}
  \caption{Mapping process between concerns, metrics and IoT devices.} \label{fig:mapping}
\end{figure*}

Based on the results achieved in the three research questions, we obtained a metric-concern mapping (RQ$1$), a metric-device mapping (RQ$2$), and feedback concerning the usability of IoT devices in a secure cyber-physical space. In \Cref{fig:mapping} we show the final mapping between concerns and IoT devices. \Cref{fig:mapping}a is the graphical representation of the concern-metric-device mapping, born from the merge of the first two research questions. For example, on one side, in the RQ$1$, the first concern (C1) has a mapping with the performance metric (M2) and the reliability metric (M4). On the other side, in the RQ$2$, the CCTV (D3) and Drone (D4) IoT devices have a mapping with the same metrics. It means that the first concern has a relation with these IoT devices. This result has been conceptualized in \Cref{fig:mapping}b. It is generated by removing the metric layer from \Cref{fig:mapping}a and connecting concerns directly with IoT devices. We grouped concerns that share the same IoT devices.


\section{Discussion} \label{sec:discussion}

\textcolor{\editcolor}{From the results and contributions outlined in the Introduction we report the following key discussion points, reflecting design principles and lessons learned from our experience.}

\textcolor{\editcolor}{\subsection{Design Principles}}
\textcolor{\editcolor}{The design principles we uncovered in the scope of our exercise essentially correspond to the metric-concern mapping that addresses RQ$1$. Specifically, the relation between a concern we reported in our experience and the metrics defined to address that concern constitute a design principle: practitioners can address the concern and track the extent to which that concern is addressed over time and during their systems' operations. We can take as example the relation between adaptability and the need to identify an anomalous trajectory (Concern 4). In the context of Concern 4 the design principle discovered indicates that knowledge over involved devices with high-level adaptability are likely to deal with the anomalous trajectories concern.} 
Furthermore, the results of the RQ$1$ indicate that it is possible to adapt non-functional metrics to other more general scenarios. For example, we discovered that 1) the adaptability metric is helpful to manage all those dynamic scenarios, as the pick-pocketing detection in a crowd scenario, 2) the usability can be used for that kind of scenario and concerns where there is an interaction between different devices that can be easily configured (e.g., monitoring of a critical scenario by sending notification to agents), 3) the reliability can be used when a scenario is not fault-proof and 4) the performance (autonomy) results in a good indicator for concerns that include many portable and wearable devices.
\newline

\textcolor{\editcolor}{\subsection{Lessons Learned}}
\textcolor{\editcolor}{Within RQ$2$ we evaluated the relationship between devices involved in our study and resulting reference architecture and metrics, extrapolating the pros and cons of each device. In the context of our work, whenever we find a relation between adaptability and smart-glassed we can assert
that in a critical scenario the wearable devices resulted to be the most appropriate choice. 
}
In the context of RQ$2$ and monitoring activity, we highlighted how smartphones resulted in being the most suitable devices.
Indeed, with an average score of 4 out of 5 smartphones appear to be easy to use, they provide an ideal battery size (compared to the other devices we took in consideration), and can be connected to a portable power bank for additional power supply.
Moreover, smartphones provide the possibility to design and implement additional software for monitoring activities. Furthermore, in the context of adaptability, smartphones can set up a communication channel with every external server (i.e., communication with the police server or with a ML server for automatic anomalous trajectory detection).
Lastly, the computational power provided by new smartphones devices offer the possibility to execute a machine-learning algorithm with an independent monitoring and prediction tool, removing external dependencies (e.g., a remote server).

With an average score of 3.75 out of 5, we have smartglasses. Like smartphones, they share the same software benefits, considering that both are equipped with Android OS. However, the battery life and drain is conspicuous, and the actual price for a pair of smartglasses makes them quite unaffordable choice.   

The CCTV has an average score of 3.5 out of 5. Assuming the Security IT Department provides access to the data could be a good option on a limited budget. Moreover, it is an appropriate choice if the monitoring and detection task's scope is focused only on specific places (e.g., a fixed camera could be enough to monitor a closed space like a stage).


The Drone scored on an average of 2.5 out of 5. If the monitoring aims to create a fault tolerance system from a software perspective, a drone may represent the most suitable option as the CCTV. Conversely, we suggest to take into consideration other devices like smartphones and smartglasses. 
Moreover, the battery drain resulted to be the most critical issue for drone devices. The tested drone could not perform more than 24 minutes seeking for a charge after that specific amount of time. 

Nonetheless, we detected different issues: 1) both smartphone and smartglasses, during the event, had some WIFI issues, having some difficulties connecting to the internal server due to the poor WIFI coverage of the event, causing some data loss; 2) Bluetooth communication between the smartphone and smart-glass were unstable. We tried to share notifications without the server's involvement, but the paring protocol between smartphone and smart-glass was unreliable, causing continuous disconnections.

\textcolor{\editcolor}{\subsection{IoT Experiences and Impacts}}
\textcolor{\editcolor}{This contribution, reflection of the feedback received on the field, finds its answer in the RQ$3$ results.}
Regarding the smartphone, all agents could complete their tasks without any problem related to the smartphone. They started and used the application provided following the scheduling without any problem. The efficiency reported was good too. We detected that the battery life was sufficient to achieve a monitoring time of at most 1h30m. Regarding the learnability, no specific issues were raised by participants. We thought that is cause they use a smartphone every day and they are practicing with them.
Differently, for the smartglasses, our participants raised many issues. 
First, they raised concerns about the overly fast battery drain. Considering the battery limitation, using it for more than 30 minutes requires an additional battery, like a power bank. Second, the device has been proved to be not easy to use for someone not necessarily used to augmented reality. Some concerns raised are the display being too small and misunderstanding about the glasses' control interface (switch and power buttons). Moreover, not all of them were able to complete the task assigned because the device suddenly crashed.

\textcolor{\editcolor}{\subsection{Platform Reference Architecture}}
\textcolor{\editcolor}{The platform architecture represents the overall contribution of this project. Expression of the Main RQ results, its main scope is to provide the best solution, in terms of IoT devices, for each concern raised by stakeholders.}
From the relations found in the RQ$1$ and RQ$2$, we defined the final mapping between concerns and device, shown in \Cref{fig:mapping}, answering to the Main RQ. From the graph, we found that some concerns can be grouped because they share the same devices.  
\textcolor{\editcolor}{More specifically, we found that smartphones and smartglasses are useful for concerns focused on anomalous trajectory and hot-spot identification (C2-C3). However, some concerns like pick-pocketing detection and real-time tracking (C4-C5) do not require any specific device, being all of them equally applicable on the field and fraudulent, criminal scenario detection in a crowd (C1) can be better solved using CCTV or drones.}

\section{Study Limitations and Threats to Validity} \label{ttv}

This work presents a threat in the experimentation process. Indeed, no specific illicit event happened during our gathering process, and no special device issues were detected. So we cannot be sure that features like \emph{Adaptability} and \emph{Robustness} have been thoroughly analyzed. Because of that, we need to run more exhaustive experimentation in different scenarios to have a more homogeneous dataset. 

Unfortunately, the data we collected is not enough to prove the multi-observation generalisability of effectiveness for our results, but it provides a valuable starting point for such generalisability. Furthermore, our data did not eventually contain any illicit activity matching the targeted scenarios, which, although it indicated a successful PaasPop event in 2019, does limit the subsequent usability of the data for some of the purposes of our study. For this reason, we are planning to participate in new social events, even and including subsequent versions of the PaasPop event, using the results obtained in this work as a starting point.

Like any study of comparable magnitude and scale, this study is affected by several threats to validity \cite{wohlin}. In what follows, we outline the major ones in our study design and execution.

\subsection{Internal Validity}

Internal validity refers to the internal consistency and structural integrity of the empirical research design.  Specifically, it refers to how many confounding factors may have been overlooked. In the scope of our study, we performed a field study in which most if not all the factors affecting the results are intrinsically uncontrollable. In this sense, however, we argue that such richness of unforeseeable enriches the validity of our contributions rather than compromising it since it reflects lessons learned from hard real-life experimentation, which was the intended aim of the study itself.

\subsection{Construct and External Validity}

From a construct and external validity perspective, our study provides several limitations. For example, while the contributions we provide may be generalizable to similar domains, we cannot say the same if the same lessons and contributions are ported to other domains. Similarly, IoT experimentation is wide and is expected to grow wider still. Therefore, we have no way of knowing whether improved sensors or other hardware may change the validity of our contributions in a few months. Nevertheless, our study, with its scope and magnitude, does represent a first attempt to report on the field study of IoT architectures in action, e.g., to figure out their limitations such that architects may take such limitations into account in their IoT design exercises. Conversely, to strengthen this side of our study and the solidity of its intended contributions, shortly, we plan to replicate the Paaspop experimentation to understand what results and lessons remain valid in subsequent replications of this same study.

\subsection{Conclusion Validity}

Conclusion validity represents the degree to which conclusions about the relationship among variables are reasonable. In the scope of the discussions of our results, we made sure to minimise possible interpretations, designing the study regarding known hypotheses. Also, our conclusions were drawn from a triangulated analysis of available data and involving triangulated stakeholders. Our study's conclusions also reflect assumptions that, although sound and reflecting the need to avoid research design mishaps, may compromise the conclusion validity. 

\section{Conclusions and Future Work} \label{sec:conclusion}
\subsection{Recap and final remarks}
In this work, we led a field study on the PaasPop event in the context of the VISOR project. Our main goal was to identify an optimal IoT architecture for cyber-physical space protection based on stakeholders' concerns and IoT devices. Consequently, we outlined three research questions that help build the architecture to relate concerns and IoT devices. First, using non-functional metrics to evaluate the relation between them and concerns (RQ$1$) or devices (RQ$2$), our scope was, in the end, to find a relation between concerns and IoT devices. Moreover, with user feedback on wearable devices (RA$3$), we want to provide an additional layer to strengthen or weaken the final decision. 

We used a card sorting approach to evaluate the first two research questions and the Krippendorff alpha coefficient to estimate the reliability of our choice. With an $\alpha$ score of $0.805$ for the RQ$1$ and $0.748$ for the RQ$2$, we can consider the relation made in these research questions trusted. 


Instead, we used five criteria from the user experience perspective: task success, task time, efficiency, learnability, and self-reported metric. We discovered that users prefer smartphones instead of smartglasses because they are more practical and know how to use them differently. However, they appreciate the wearable technology of smart-glass, avoiding holding the phone for the entire monitoring activity.

Lastly, we create a mapping between concerns and IoT devices in the main research question, cross-referencing information from RQ$1$ and RQ$2$.

\subsection{Practical Contributions and Replication Package}
As an outcome of this work, we created a prototype of a graphical user interface for visualizing in real-time all the IoT devices involved and used in this study. Furthermore, we collected video and frames for our future experimentations. The prototype, the data, and the dataset discussed in \Cref{sec:dataset} can be reached by the following GitHub repository link: https://github.com/jade-lab/Paaspop2019-Replication-Package

\subsection{Future Work} \label{sec:future_works} 

In the scope of this work, we led a preliminary field study designed to create an infrastructure to relate concern and IoT devices. Stemming from our lessons learned and contributions, the next step is, starting from the technologies involved in this study, to scale up the proposed architecture, e.g., to detect and prevent multiple illicit actions within specific contingency plans, and with more triangulated data and more devices. Similarly, in the future, we plan to replicate the study in other domains and, as aforementioned, at a more significant or varying scale. Finally, the research intends to strengthen the contribution's generalizability and determine the limitations of the proposed contributions' applicability. \textcolor{\editcolor}{Finally, our results do not warrant any considerations beyond the scope of our experiences and the report we produced in this manuscript. Conversely, we are planning to further evaluate our proposed results and experiences with more replications of the same experiment we reported here, adding on top of any modelling exercise the use of sound mathematical modelling and other more formal approaches which may strengthen the external validity of the findings reported in this paper.}

\section*{Acknowledgements}
The work is supported by the EU H2020  framework programme, grant ``ANITA'' under Grant No.: 787061 and grant ``PRoTECT'' under Grant No.: 815356.

\bibliography{main.bib}

\end{document}